\documentclass[11pt]{article}

\usepackage[margin=1in]{geometry}
\usepackage{graphicx}
\usepackage{amsmath,amssymb}
\usepackage[hidelinks]{hyperref}
\usepackage{placeins}
\usepackage[all]{hypcap}
\usepackage{chngcntr}

\counterwithout{table}{section}
\counterwithout{figure}{section}

\title{\textbf{The Impact of Lesion Focus on the Performance of AI-Based Melanoma Classification}}

\author{
Tanay Donde\\
Department of Computer Science\\
University of Illinois Urbana-Champaign\\
\texttt{tanaydonde@gmail.com}\\[0.3em]
}

\date{} % leave blank

\begin{document}
\maketitle

\begin{abstract}
Melanoma is the most lethal subtype of skin cancer, and early and accurate detection of this disease can greatly improve patients' outcomes~\cite{skincancerstats}. Although machine learning models, especially convolutional neural networks (CNNs), have shown great potential in automating melanoma classification, their diagnostic reliability still suffers due to inconsistent focus on lesion areas~\cite{ker2018deep}. In this study, we analyze the relationship between lesion attention and diagnostic performance, involving masked images, bounding box detection, and transfer learning. We used multiple explainability and sensitivity analysis approaches to investigate how well models aligned their attention with lesion areas and how this alignment correlated with precision, recall, and F1-score. Results showed that models with a higher focus on lesion areas achieved better diagnostic performance, suggesting the potential of interpretable AI in medical diagnostics. This study provides a foundation for developing more accurate and trustworthy melanoma classification models in the future.
\end{abstract}

\section{Introduction}

Skin cancer is the most common form of cancer worldwide, and melanoma is its most lethal variant. Early and accurate detection is critical for improving patient outcomes. When melanoma is diagnosed at an early stage, the five-year survival rate exceeds 99\%. However, this rate drops to approximately 32\% once the cancer has spread to distant organs~\cite{skincancerstats}. 

Deep learning models, particularly convolutional neural networks (CNNs), have shown strong potential in automating medical image analysis~\cite{ker2018deep}. Despite this promise, many models lack reliability due to their ``black box'' nature, as they do not provide clear explanations for their predictions~\cite{rudin2019stop}. In the context of melanoma diagnosis, a key factor underlying reliable models is lesion attention---the extent to which a model focuses on lesion regions within dermoscopic images.

\begin{figure}[t]
    \centering
    \includegraphics[width=\linewidth]{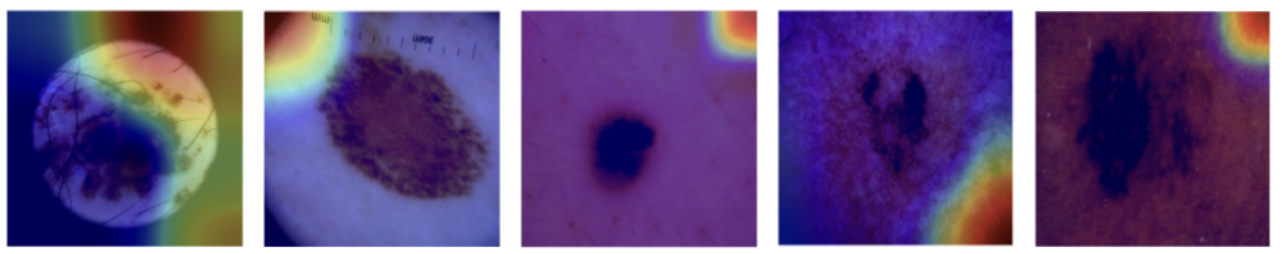}
    \caption{Grad-CAM results from the baseline InceptionV3 model, demonstrating misaligned attention away from the lesion area across various examples during melanoma classification.}
    \label{fig:gradcam_baseline}
\end{figure}

Using Grad-CAM, a widely used explainability method, we observed that baseline CNN models often fail to attend to lesion areas, instead focusing on irrelevant background regions~\cite{selvaraju2019gradcam}. Figure~\ref{fig:gradcam_baseline} illustrates several examples where model attention is misaligned with the lesion during melanoma classification.

\FloatBarrier

\begin{figure}[htbp]
    \centering
    \includegraphics[width=0.3\linewidth]{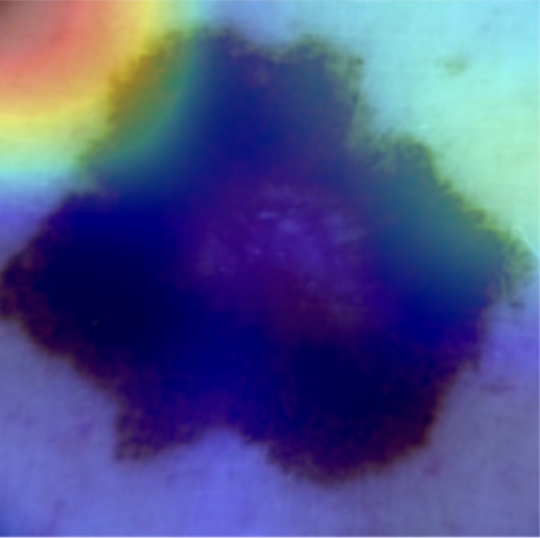}
    \caption{Grad-CAM explanation of a correctly predicted melanoma case where the model’s attention is misaligned, focusing away from the lesion.}
    \label{fig:gradcam_misaligned}
\end{figure}

Additionally, we observed cases where models correctly predicted melanoma while exhibiting attention maps that were almost completely misaligned with the lesion. Figure~\ref{fig:gradcam_misaligned} shows an example of a correctly classified melanoma case in which the model’s attention is focused away from the region of interest.

This brings us to our central question: does increased model attention on lesion areas correspond to better diagnostic performance? Intuitively, this seems like an obvious question. Increasing a model’s attention to the lesion area should naturally improve diagnostic accuracy. However, larger models often base their decisions on features or contextual cues unrelated to the task, making this relationship less straightforward. For instance, in melanoma classification, models can learn to associate background features, such as skin color, with specific classes rather than focusing on the actual lesion. 

Answering this question is critical for improving both the accuracy and transparency of AI-driven melanoma classification models. If increased lesion focus does correlate with better performance, it could provide a pathway toward designing models that are both accurate and reliable in medical settings. By consistently focusing on lesion areas, these models increase interpretability, addressing concerns about the “black box” nature of AI in medical imaging. Additionally, these findings could guide the development of explainability-driven training techniques, allowing models to learn diagnostically relevant features more effectively.

To explore this hypothesis, we designed experiments to evaluate whether improved attention to lesion areas enhances diagnostic performance. These experiments included bounding box detection, pretraining, and transfer learning using masked image datasets, where lesion areas were isolated. Alongside these experiments, we employed multiple explainability and sensitivity analysis methods, including Grad-CAM, Sobol', and RISE (Randomized Input Sampling for Explanation), to qualitatively assess models’ focus to lesion areas~\cite{selvaraju2019gradcam,sobol2021variance,petsiuk2018rise}. Together, these approaches provided a comprehensive framework for assessing how model alignment with lesion areas influences diagnostic accuracy and offered insights into the development of more interpretable and diagnostically relevant AI models.

\section{Related Work}

Several studies have explored various applications of AI models for skin lesion analysis, including lesion detection, localization, and classification. While significant progress has been made in these areas, there remains a gap in understanding whether improving lesion focus specifically correlates with enhanced diagnostic performance. Below, we examine related work to contextualize our study and demonstrate its unique contribution to AI-driven skin lesion analysis.

Adegun and Viriri provided a comprehensive review of state-of-the-art techniques, discussing pre-processing, segmentation, and classification methods for skin lesion analysis~\cite{adegun2021survey}. While their review included segmentation techniques as part of the workflow for lesion analysis, it did not experimentally test how segmentation impacts diagnostic accuracy, nor did it compare segmentation-based approaches to those without segmentation. In contrast, our study directly investigates this relationship, providing novel insights into the role of lesion segmentation in melanoma classification.

Efforts to integrate lesion localization into melanoma detection have shown potential. Taghizadeh and Mohammadi proposed a two-step pipeline using YOLOv3 for lesion detection and SegNet for segmentation, achieving high localization accuracy~\cite{taghizadeh2023yolo}. While their work demonstrated the feasibility of lesion localization, it focused exclusively on segmentation tasks and did not evaluate the impact of lesion focus on diagnostic accuracy for melanoma classification.

Recent work by Getamesay Haile Dagnaw, Meryam El Mouhtadi, and Musa Mustapha explored the integration of vision transformers (ViTs) and convolutional neural networks (CNNs) for skin cancer classification, incorporating explainability methods to interpret model decisions~\cite{dagnaw2024vit}. Their study demonstrated the utility of explainable AI in identifying relevant features for clinical diagnostics, reinforcing the importance of transparency in AI-driven healthcare solutions. While this approach aligns with our emphasis on model interpretability, our study is distinct in its focus on the relationship between lesion attention and diagnostic performance. Additionally, our work investigated specific techniques, such as masked image datasets, bounding box detection, and transfer learning, for enhancing lesion focus, providing a more detailed analysis of model attention alignment. These differences highlight the unique contribution of our study, addressing a distinct and less-explored aspect of AI-driven melanoma classification.

Despite these advancements, there remains limited evidence on whether improving lesion focus correlates with enhanced diagnostic accuracy in dermoscopic imaging. To address this gap, our study investigates this relationship by employing multiple explainability methods, to assess model attention. This research contributes to the broader effort to make AI-driven melanoma classification more interpretable and reliable by offering a deeper understanding of model reasoning.

\FloatBarrier

\section{Methods}

To investigate whether increased attention to the lesion area correlates with improvements in diagnostic accuracy, we conducted a series of experiments. These experiments explored various techniques to enhance the model’s attention on the lesion, such as segmentation-based approaches and object detection tasks, with the ultimate goal of understanding the relationship between lesion focus and performance. Specific details of the experiments are provided in the following sections. For all experiments, multiple hyperparameter configurations were tested, and the best-performing models were selected based on F1-score. Grad-CAM, Sobol’, and RISE were used only to compare the final models from each experiment with the baseline model to assess lesion focus qualitatively. All experiments were implemented using Keras and TensorFlow on Kaggle’s platform, leveraging GPU acceleration where available~\cite{chollet2015keras,abadi2016tensorflow}. All input images were normalized to the range $[-1, 1]$ during preprocessing to improve numerical stability and training efficiency. 

\subsection{Datasets}

This study utilized two publicly available datasets, ISIC-2019 and HAM10000, for all experiments~\cite{codella2018isic,tschandl2018ham10000,hernandez2024bcn20000}.

The ISIC-2019 dataset is an aggregate dataset comprising 25,331 dermoscopic images sourced from various sub-datasets, including HAM10000. The 25,331 images were split into 75\% training (18,999 images) and 25\% validation (6,332 images) for the baseline model. Validation images were selected randomly. To address the imbalance between melanoma (mel) and non-melanoma (non-mel) classes, oversampling was applied to the training set in all experiments to ensure equal representation of both classes.

The HAM10000 dataset, which contributes to ISIC-2019, contains 10,015 dermoscopic images and diagnostic labels. While the original dataset does not include segmentation masks, a Kaggle dataset derived from HAM10000 provides these masks~\cite{tschandl2020ham10000seg}. They were initially generated using a Fully Convolutional Network (FCN) and later reviewed and corrected by a dermatologist using FIJI's free-hand selection tool. These curated masks were utilized in this study for segmentation-based experiments and for generating bounding boxes.

Both HAM10000 and ISIC-2019 contain multiple classes of skin lesions, as shown in Table~\ref{tab:class_distribution}~\cite{hoang2022multiclass}.

\begin{table}[htbp]
\centering
\begin{tabular}{lcc}
\hline
\textbf{Class Name} & \textbf{HAM10000 Number of Images} & \textbf{ISIC-2019 Number of Images} \\
\hline
AKIEC & 327 & 867 \\
BCC   & 514 & 3,323 \\
BKL   & 1,099 & 2,624 \\
DF    & 115 & 239 \\
MEL   & 1,113 & 4,522 \\
NV    & 6,705 & 12,875 \\
VASC  & 142 & 628 \\
SCC   & -- & 253 \\
\hline
\textbf{Total} & \textbf{10,015} & \textbf{25,331} \\
\hline
\end{tabular}
\caption{Class distribution of skin lesion images in the HAM10000 and ISIC-2019 datasets, showing the number of images available for each lesion type.}
\label{tab:class_distribution}
\end{table}

\begin{figure}[htbp]
    \centering
    \includegraphics[width=\linewidth]{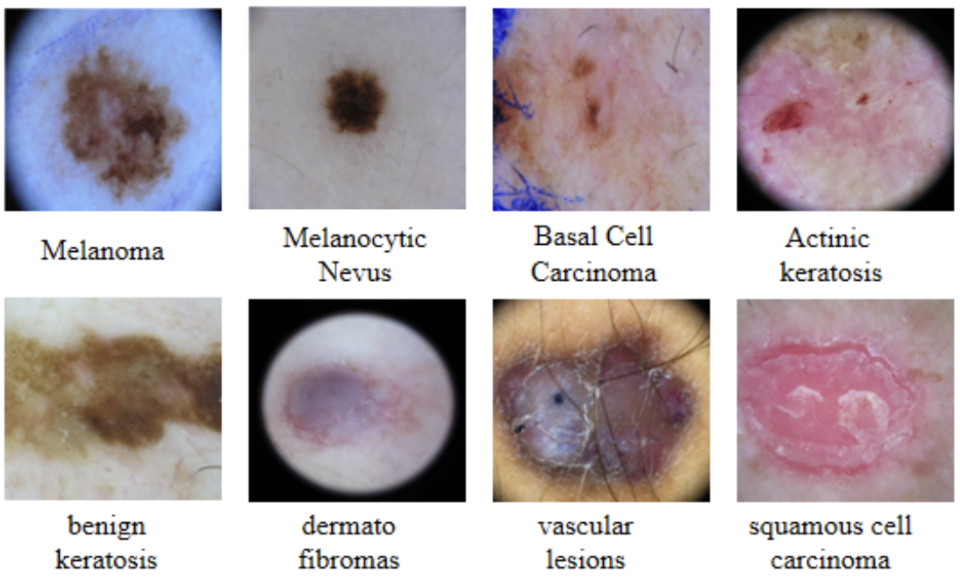}
    \caption{Representative images from each class in the ISIC-2019 dataset.}
    \label{fig:isic2019_classes}
\end{figure}

HAM10000 includes seven classes: Actinic Keratoses and Intraepithelial Carcinoma (AKIEC), Basal Cell Carcinoma (BCC), Benign Keratosis-like Lesions (BKL), Dermatofibroma (DF), Melanoma (MEL), Melanocytic Nevi (NV), and Vascular Lesions (VASC). ISIC-2019 expands on this by including one additional class, Squamous Cell Carcinoma (SCC). A snippet of the ISIC-2019 dataset is provided in Figure~\ref{fig:isic2019_classes}~\cite{li2024semisupervised}.

For this study, as our focus was on binary classification, these classes were grouped into melanoma and non-melanoma categories.

\subsection{Evaluation Metrics}

Model performance was evaluated using a combination of quantitative and qualitative metrics:

\begin{itemize}
    \item \textbf{Precision:} Proportion of true positive predictions among all positive predictions, measuring the model’s ability to avoid false positives.
    \item \textbf{Recall:} Proportion of true positive predictions among all actual positive cases, reflecting sensitivity to positive samples.
    \item \textbf{F1-score:} Harmonic mean of precision and recall. This metric was used to select the best-performing models in each experiment due to class imbalance.
    \item \textbf{Explainability methods:} Grad-CAM, Sobol’, and RISE were used to qualitatively assess alignment between model attention and lesion regions. These methods were applied only to the best-performing models from each experiment and compared against the baseline.
\end{itemize}

Heatmaps were generated using the \texttt{Xplique} library, which provided implementations for each method ~\cite{fel2022xplique}:
\begin{itemize}
    \item \texttt{xplique.attributions.GradCAM} for Grad-CAM
    \item \texttt{xplique.attributions.SobolAttributionMethod} for Sobol’
    \item \texttt{xplique.attributions.Rise} for RISE
\end{itemize}

The resulting heatmaps were visualized by overlaying them on the input images using a jet colormap with an opacity of $\alpha = 0.5$, ensuring that both the heatmap and the underlying lesion remained visible.

\begin{figure}[t]
    \centering
    \includegraphics[width=\linewidth]{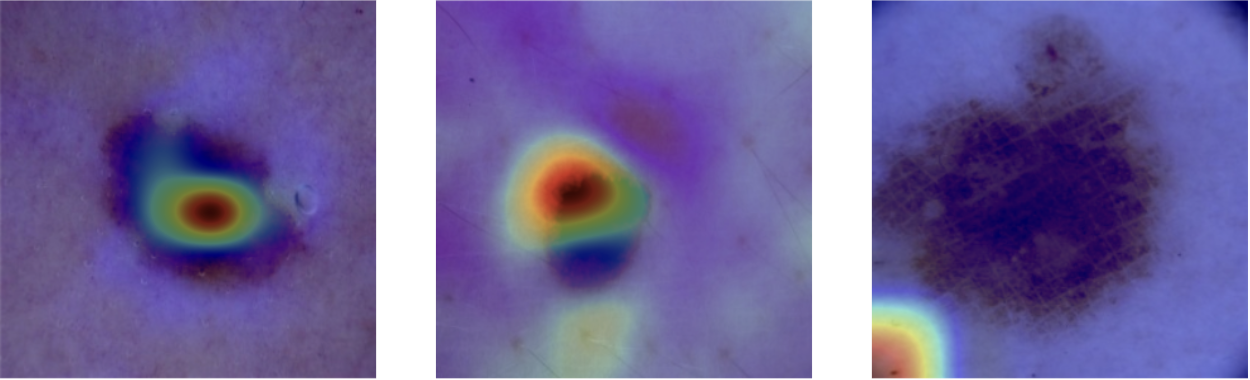}
    \caption{Visual examples illustrating the three attention alignment categories: fully aligned (left), partially aligned (center), and misaligned (right).}
    \label{fig:attention_categories}
\end{figure}

Assessing lesion focus using explainability heatmaps can involve ambiguous interpretations. The notion of being ``on the lesion'' is inherently imprecise, as attention may overlap the lesion to varying degrees or highlight unrelated regions. Due to dataset constraints and preprocessing differences, a unified quantitative attention--mask overlap metric was not applicable across all models. We therefore relied on controlled qualitative analysis and defined clear, consistent categories for visual assessment (see Figure~\ref{fig:attention_categories}):

\begin{itemize}
    \item \textbf{Fully aligned:} The heatmap is primarily concentrated on the lesion, with no significant attention elsewhere.
    \item \textbf{Partially aligned:} The heatmap overlaps the lesion but also extends noticeably beyond its boundaries or highlights unrelated regions.
    \item \textbf{Misaligned:} The heatmap minimally overlaps or does not overlap the lesion, with attention concentrated outside the lesion area.
\end{itemize}

To ensure consistency in the qualitative evaluation, all visual assessments were performed by the same individual according to these predefined categories.

\subsection{Baseline Model}

To select an appropriate baseline architecture, we evaluated three widely used models: ResNet50~\cite{he2016deep}, BEiT v2~\cite{peng2022beitv2}, and InceptionV3~\cite{szegedy2016rethinking}. 
ResNet50 is a convolutional neural network that leverages residual connections to enable effective training of very deep models. 
BEiT v2 is a transformer-based architecture designed for large-scale image modeling but is computationally intensive. 
InceptionV3 is a convolutional neural network that efficiently captures multi-scale features through parallel convolutional pathways. For this study, we used the InceptionV3 and ResNet50 implementations provided by \texttt{tensorflow.keras.applications}~\cite{abadi2016tensorflow}. 
The BEiT v2 model was obtained from the \texttt{keras\_cv\_attention\_models} library~\cite{keras_cv_attention_models}.

Each model was trained and evaluated over multiple iterations on the ISIC-2019 dataset for binary classification of melanoma versus non-melanoma. ResNet50 achieved a best F1-score of 0.661, while BEiT v2 and InceptionV3 achieved best F1-scores of 0.716 and 0.718, respectively. 
Although BEiT v2 performed comparably to InceptionV3, its substantially longer training time made it less practical for the extensive experiments conducted in this study. Consequently, InceptionV3 was selected as the baseline architecture due to its balance of strong performance and training efficiency.

The baseline InceptionV3 model was initialized with pretrained ImageNet weights and fine-tuned using the Adamax optimizer~\cite{kingma2017adam}. 
An exponential learning rate decay schedule was employed, along with a validation-based learning rate callback to dynamically adjust the learning rate during training. 

This baseline model served as the control for subsequent experiments involving alternative dataset configurations and training strategies.

\subsection{YOLO Model}

In this experiment, we investigated whether incorporating object detection prior to classification could improve diagnostic performance by enabling the model to localize the lesion before prediction. 
To this end, we employed the YOLO (You Only Look Once) framework, a real-time object detection architecture designed to jointly perform object localization and classification within a single forward pass~\cite{redmon2016yolo}.

\begin{figure}[htbp]
    \centering
    \includegraphics[width=\linewidth]{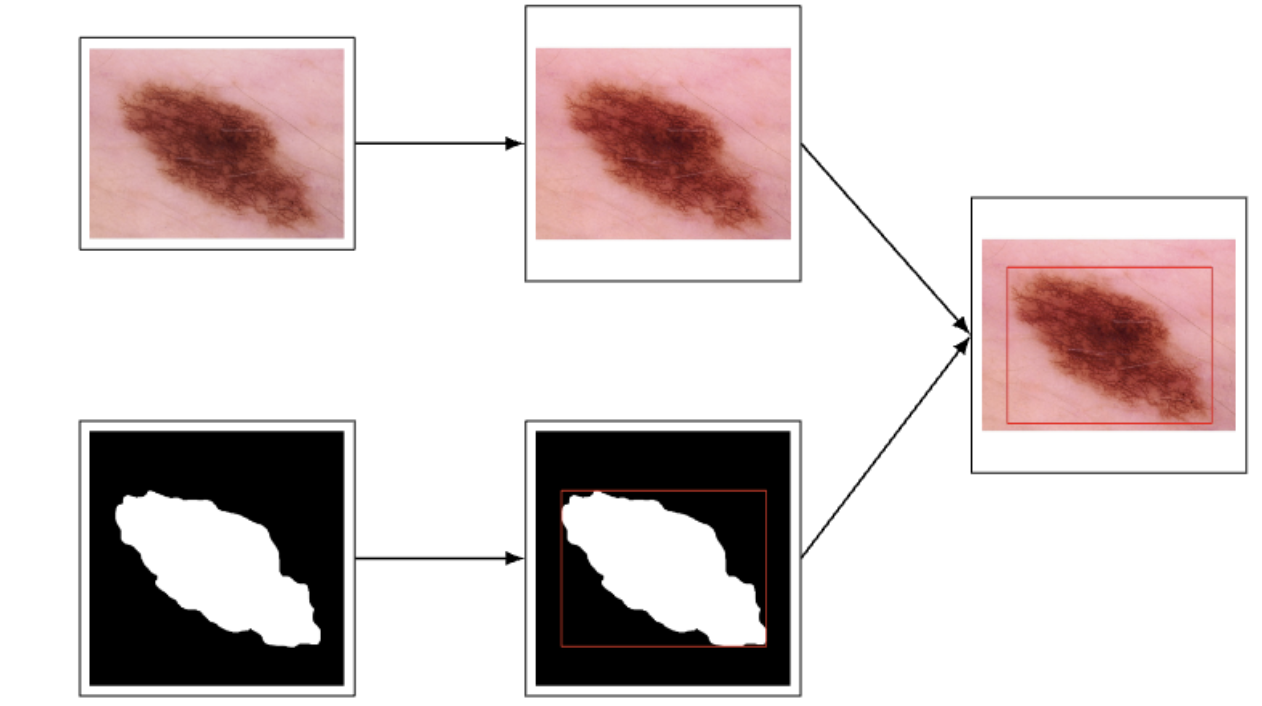}
    \caption{Flowchart illustrating the process of generating bounding boxes from lesion segmentation masks in the HAM10000 dataset.}
    \label{fig:yolo_bbox_generation}
\end{figure}

The YOLO model was trained and evaluated on the HAM10000 dataset for lesion detection and classification. The dataset was split into 80\% training and 20\% validation. Bounding boxes were generated prior to training to adapt the dataset for object detection. First, each lesion image was padded to maintain consistent spatial dimensions. Next, the corresponding segmentation mask provided by the HAM10000 dataset was used to determine the tight bounding box enclosing the lesion. Finally, this bounding box was applied to the padded image to produce the detection target. Figure~\ref{fig:yolo_bbox_generation} illustrates this bounding box generation process.

\FloatBarrier

For implementation, we utilized the Ultralytics YOLOv8 framework, selected for its robustness and efficiency in object detection tasks~\cite{jocher2023ultralytics}.

\subsection{Masked Image Experiments}

During the masked image experiments, we investigated whether explicitly increasing model focus on the lesion region would correlate with improvements in diagnostic performance. To this end, masked images were generated in which the lesion was isolated and background information was removed.

\begin{figure}[t]
    \centering
    \includegraphics[width=\linewidth]{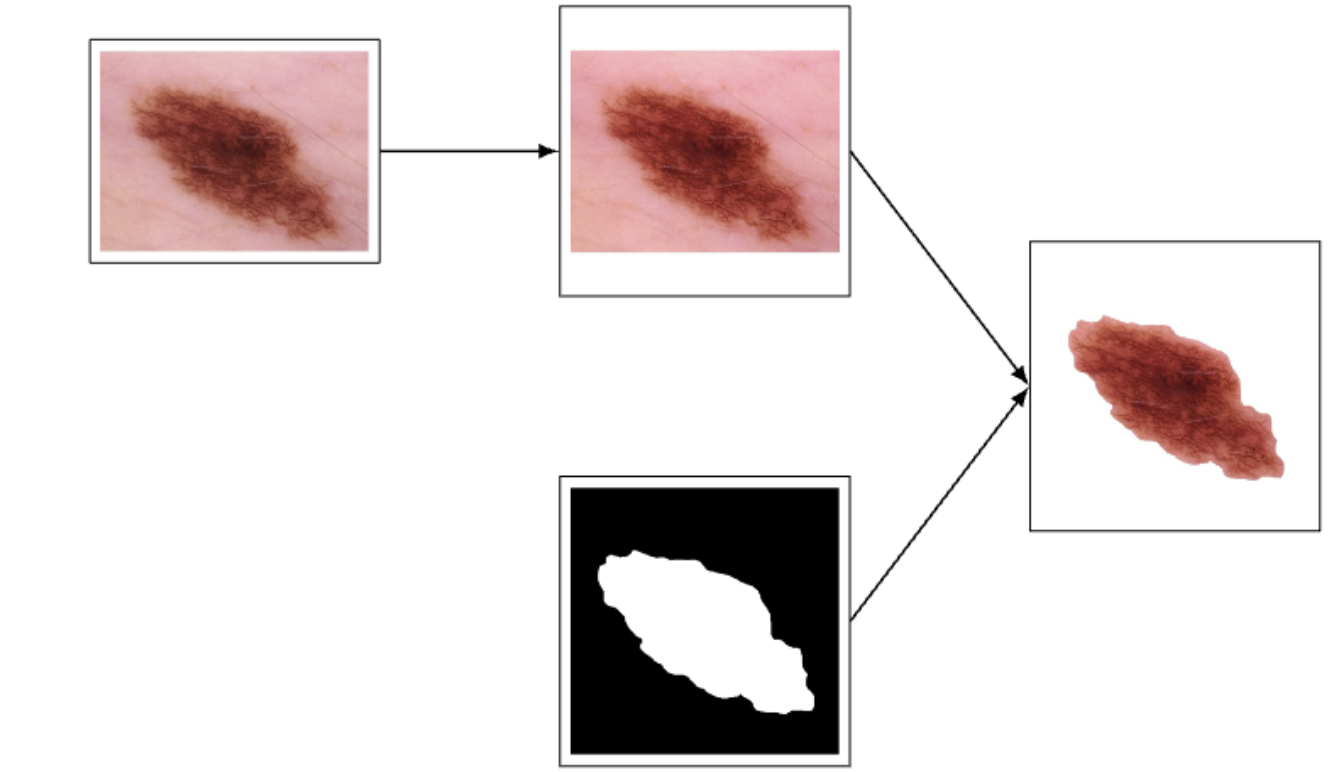}
    \caption{Flowchart illustrating the process of generating masked images for lesion segmentation experiments.}
    \label{fig:masked_image_generation}
\end{figure}

The process began with the original lesion image, which was padded to ensure consistent spatial dimensions. Next, the corresponding segmentation mask was applied to isolate the lesion region. All pixels outside the lesion were replaced with a uniform white background, resulting in the final masked image. Figure~\ref{fig:masked_image_generation} illustrates this step-by-step masking procedure.

\FloatBarrier

\subsubsection{Masked-to-Regular Transfer}

To investigate the impact of lesion segmentation, a new InceptionV3 model, initialized with pretrained ImageNet weights, was trained only on masked images from HAM10000, where the lesion was isolated, and the background was replaced with white. The dataset was split into 80\% training and 20\% validation. During this process, multiple learning rates and hyperparameter configurations were tested to identify the best-performing model. Once the best-performing configuration was identified, that model’s weights were transferred to the regular ISIC-2019 dataset for further training. In this second phase, multiple learning rates and hyperparameter configurations were again explored, and the best-performing setup was selected for the final model. Various layers were frozen during transfer to retain knowledge gained from the masked images. We ensured the ISIC-2019 validation set did not contain any images that were part of the HAM10000 masked training set. This step was necessary to ensure fair evaluation, as testing on images previously used for training would not provide an accurate measure of the model’s performance. These models were evaluated using F1-score.

Explainability methods were not used during model training or hyperparameter tuning. Instead, they were applied after identifying the final best-performing model, comparing its focus on lesion areas to the original baseline model qualitatively.

\subsubsection{Combined Masked and Regular Datasets}

To evaluate the combined effect of masked and regular images, a dataset consisting of 15{,}316 regular images and 10{,}015 masked images was constructed. Models were trained using different proportions of regular and masked images, specifically 50\%, 75\%, and 90\% regular images, with the remaining portion composed of masked images. Two training strategies were evaluated:

\begin{itemize}
    \item \textbf{Method 1:} The model was initialized with ImageNet-pretrained InceptionV3 weights and trained for 30 epochs on the combined datasets. 
    Multiple hyperparameter configurations, including different learning rates, were tested, and the best-performing model was selected based on F1-score.

    \item \textbf{Method 2:} The baseline InceptionV3 model trained on the regular ISIC-2019 dataset was used as the starting point. 
    This model was subsequently fine-tuned for 15 epochs on the combined datasets. 
    As in Method~1, multiple hyperparameter configurations were explored, and the best configuration was selected based on performance metrics.
\end{itemize}

For Method~2, care was taken to ensure that the validation set did not contain any images that had been included in the training set of the baseline InceptionV3 model.

Following training, the best-performing model across both methods, as determined by F1-score, was selected for further analysis. Grad-CAM, Sobol’, and RISE were applied exclusively to this model to assess attention focus, and the results were compared against those obtained from the original baseline model.

\section{Results and Discussion}

This section summarizes the performance of various models trained and evaluated during the study.

\subsection{Baseline Model Performance}

The baseline InceptionV3 model was trained for 15 epochs on the ISIC-2019 dataset, achieving an accuracy of 90.22\% and an F1-score of 0.743 (precision: 0.8096, recall: 0.6868). Grad-CAM, Sobol’, and RISE were employed to analyze the model’s attention on lesion regions for a sample of 40 images, consisting of 20 true melanoma and 20 true non-melanoma cases. This stratified sampling enabled a balanced analysis of attention alignment across true class labels. The results are summarized in Table~\ref{tab:attention_alignment_baseline}.

\begin{table}[htbp]
\centering
\begin{tabular}{llccc}
\hline
\textbf{Explainability Method} & \textbf{Alignment} & \textbf{True Melanoma} & \textbf{True Non-Melanoma} & \textbf{Total} \\
\hline
\textbf{Grad-CAM} & Fully Aligned      & 4  & 0  & 4  \\
                  & Partially Aligned  & 7  & 12 & 19 \\
                  & Misaligned         & 9  & 8  & 17 \\
\hline
\textbf{Sobol'}   & Fully Aligned      & 10 & 3  & 13 \\
                  & Partially Aligned  & 10 & 13 & 23 \\
                  & Misaligned         & 0  & 4  & 4  \\
\hline
\textbf{RISE}     & Fully Aligned      & 13 & 2  & 15 \\
                  & Partially Aligned  & 6  & 8  & 14 \\
                  & Misaligned         & 1  & 10 & 11 \\
\hline
\end{tabular}
\caption{Alignment of Grad-CAM, Sobol’, and RISE attention maps generated by the baseline InceptionV3 model for a balanced sample of 40 images (20 true melanoma and 20 true non-melanoma). Alignment categories include Fully Aligned, Partially Aligned, and Misaligned. Counts indicate the number of images in each category for the corresponding explainability method and true class label.}
\label{tab:attention_alignment_baseline}
\end{table}

\FloatBarrier

The significant variation in results across Grad-CAM, Sobol’, and RISE can be attributed to the distinct mechanisms underlying each explainability method. Grad-CAM, being a gradient-based approach, focuses primarily on the most salient regions influencing the model's output. In contrast, Sobol’ employs sensitivity analysis to evaluate the contribution of different input regions, which may provide a broader perspective but lack fine-grained localization. RISE, on the other hand, generates heatmaps using random perturbations, which can capture global patterns but might dilute attention on specific lesion areas. These methodological differences explain the varying levels of alignment observed across the three techniques.

\subsection{YOLO Model Performance}

The YOLO model was trained on the HAM10000 dataset to detect lesion locations and classify them as melanoma or non-melanoma. The model achieved an accuracy of 85.13\% but a notably low F1-score of 0.348. While the high accuracy initially suggests strong performance, the divergence from the F1-score indicates that the model heavily biased its predictions toward the majority class (non-melanoma), effectively failing to identify positive melanoma cases. Since the InceptionV3 baseline achieved a much higher F1-score (0.743) on the exact same dataset, the failure cannot be attributed to class imbalance alone. Instead, these results suggest that YOLO’s architecture struggled to capture the fine-grained textural distinctions required for melanoma diagnosis, which standard CNNs like InceptionV3 preserved. Given that the model failed to learn a robust discriminative signal, further interpretability analyses (Grad-CAM, Sobol’, and RISE) were not conducted for the YOLO architecture.

\subsection{Masked Image Experiment Results}

\subsubsection{Masked-to-Regular Transfer Results}

The best-performing model, initially trained exclusively on masked images, was subsequently fine-tuned on the regular ISIC-2019 dataset, achieving an accuracy of 90.19\% and an F1-score of 0.734 (precision: 0.831, recall: 0.657). The strongest performance was obtained by unfreezing only the first 200 layers and the final few layers of the network during transfer learning. Compared to the baseline model, this approach resulted in slightly lower F1-score and recall but yielded a modest improvement in precision.

To compare lesion attention with the baseline model, a stratified sample of 40 images was selected, consisting of 20 true melanoma and 20 true non-melanoma cases. Grad-CAM, Sobol’, and RISE were applied to qualitatively assess the alignment of model attention with lesion regions. This analysis enabled evaluation of how fine-tuning influenced lesion focus relative to the baseline model. The results are summarized in Table~\ref{tab:attention_alignment_masked_to_regular}.

\begin{table}[htbp]
\centering
\begin{tabular}{llccc}
\hline
\textbf{Explainability Method} & \textbf{Alignment} & \textbf{True Melanoma} & \textbf{True Non-Melanoma} & \textbf{Total} \\
\hline
\textbf{Grad-CAM} & Fully Aligned      & 3  & 6  & 9  \\
                  & Partially Aligned  & 5  & 11 & 16 \\
                  & Misaligned         & 12 & 3  & 15 \\
\hline
\textbf{Sobol'}   & Fully Aligned      & 11 & 4  & 15 \\
                  & Partially Aligned  & 8  & 13 & 21 \\
                  & Misaligned         & 1  & 3  & 4  \\
\hline
\textbf{RISE}     & Fully Aligned      & 5  & 4  & 9  \\
                  & Partially Aligned  & 13 & 10 & 23 \\
                  & Misaligned         & 2  & 6  & 8  \\
\hline
\end{tabular}
\caption{Alignment of Grad-CAM, Sobol’, and RISE attention maps generated by the masked-to-regular model for a balanced sample of 40 images (20 true melanoma and 20 true non-melanoma). The interpretation of values follows the same format as Table~\ref{tab:attention_alignment_baseline}.}
\label{tab:attention_alignment_masked_to_regular}
\end{table}

For true melanoma cases, the results show a clear correlation between degraded feature focus and reduced diagnostic sensitivity. The masked-to-regular model exhibited increased attention misalignment across all explainability methods (most notably in Grad-CAM, where misaligned instances rose from 9 to 12). This loss of focus directly tracked with a drop in Recall (from 0.687 to 0.657), confirming that when the model fails to attend to diagnostic lesion features, it fails to correctly identify the pathology, resulting in missed diagnoses (false negatives).

Conversely, true non-melanoma cases demonstrated a correlation between sharper focus and improved predictive precision. The model showed significantly better alignment for these images, with misaligned instances dropping across Grad-CAM (8 to 3), Sobol’ (4 to 3), and RISE (10 to 6). This tighter focus on the actual lesion allowed the model to correctly reject benign cases rather than being confused by background noise, reducing False Positives. Consequently, this reduction in errors drove the increase in Precision from 0.810 to 0.831.

\subsubsection{Combined Masked and Regular Dataset Results}

\paragraph{Training with ImageNet Pretrained Weights}\mbox{}

In this method, a new InceptionV3 model, initialized with ImageNet-pretrained weights, was trained on a combined dataset of masked and regular images for 30 epochs. Three dataset compositions were tested, containing 50\%, 75\%, and 90\% of regular images, with the remainder consisting of masked images. The key results, including accuracy, precision, recall, and F1-score, are summarized in Table~\ref{tab:combined_method1_results}.

\begin{table}[htbp]
\centering
\begin{tabular}{lcccc}
\hline
\textbf{Percent of Regular Images} & \textbf{Accuracy (\%)} & \textbf{Precision (\%)} & \textbf{Recall (\%)} & \textbf{F1-score} \\
\hline
50 & 87.12 & 83.80 & 45.32 & 0.588 \\
75 & 90.19 & 87.97 & 59.67 & 0.711 \\
90 & 90.31 & 84.48 & 64.96 & 0.734 \\
\hline
\end{tabular}
\caption{Performance metrics for Method~1, showing accuracy, precision, recall, and F1-score across different proportions of regular images in the combined dataset.}
\label{tab:combined_method1_results}
\end{table}

\paragraph{Training With Skin-Cancer Specific Pretrained Weights}\mbox{}

In this method, we used the baseline InceptionV3 model pretrained on the ISIC-2019 regular dataset as the starting point. The model was fine-tuned on a combined dataset of masked and regular images. The dataset compositions tested included 50\%, 75\%, and 90\% regular images, with the remainder consisting of masked images. The key results, including accuracy, precision, recall, and F1-score, are summarized in Table~\ref{tab:combined_method2_results}.

\begin{table}[htbp]
\centering
\begin{tabular}{lcccc}
\hline
\textbf{Percent of Regular Images} & \textbf{Accuracy (\%)} & \textbf{Precision (\%)} & \textbf{Recall (\%)} & \textbf{F1-score} \\
\hline
50 & 90.78 & 86.84 & 54.98 & 0.673 \\
\textbf{75} & \textbf{91.87} & \textbf{85.30} & \textbf{71.90} & \textbf{0.780} \\
90 & 91.09 & 94.03 & 51.22 & 0.663 \\
\hline
\end{tabular}
\caption{Performance metrics for Method~1, showing accuracy, precision, recall, and F1-score across different proportions of regular images in the combined dataset.}
\label{tab:combined_method2_results}
\end{table}

\paragraph{Combined Dataset Best-Performing Model Analysis}\mbox{}

The best model was selected by comparing the F1-scores of the top-performing models from Method 1 and Method 2. The 75\% regular image model from Method 2 (bolded in Table 5) was chosen as the best-performing model. This model achieved an accuracy of 91.87\% and an F1-score of 0.780 (with precision of 0.853 and recall of 0.719), outperforming the baseline model across all quantitative metrics (precision, recall, and f1-score).

To evaluate the best-performing model’s attention alignment, a sample of 40 images (20 melanoma and 20 non-melanoma) was analyzed using Grad-CAM, Sobol’, and RISE explainability methods. The results of this qualitative assessment are summarized in Table~\ref{tab:attention_alignment_combined}.

\begin{table}[htbp]
\centering
\begin{tabular}{llccc}
\hline
\textbf{Explainability Method} & \textbf{Alignment} & \textbf{True Melanoma} & \textbf{True Non-Melanoma} & \textbf{Total} \\
\hline
\textbf{Grad-CAM} & Fully Aligned     & 7  & 1  & 8  \\
                  & Partially Aligned & 3  & 12 & 15 \\
                  & Misaligned        & 10 & 7  & 17 \\
\hline
\textbf{Sobol'}   & Fully Aligned     & 15 & 7  & 22 \\
                  & Partially Aligned & 5  & 12 & 17 \\
                  & Misaligned        & 0  & 1  & 1  \\
\hline
\textbf{RISE}     & Fully Aligned     & 12 & 7  & 19 \\
                  & Partially Aligned & 5  & 8  & 13 \\
                  & Misaligned        & 3  & 5  & 8  \\
\hline
\end{tabular}
\caption{Alignment of Grad-CAM, Sobol’, and RISE attention maps generated by the combined dataset model for a balanced sample of 40 images (20 true melanoma and 20 true non-melanoma). The interpretation of values follows the same format as described for Table~\ref{tab:attention_alignment_baseline}.}
\label{tab:attention_alignment_combined}
\end{table}

\FloatBarrier

For true melanoma cases, the combined dataset model demonstrated a positive correlation between improved attention alignment and diagnostic sensitivity. The model exhibited increased alignment across two of the three explainability methods, with fully aligned instances rising for Grad-CAM (from 4 to 7) and Sobol’ (from 10 to 15). While there was a marginal increase in misaligned instances for Grad-CAM and RISE, the substantial gain in fully aligned attention maps indicates a net improvement in the model's ability to localize lesion features. This overall enhancement in focus corresponds directly with the observed increase in Recall (from 0.6868 to 0.719), confirming that better attention to relevant pathology contributed to the model's ability to classify true melanoma cases.

Additionally, true non-melanoma cases exhibited a comprehensive correlation between sharper focus and improved predictive precision. The model showed notable improvements in alignment across all explainability methods, evidenced by increases in fully aligned instances for Grad-CAM (0 to 1), Sobol’ (3 to 7), and RISE (2 to 7). Furthermore, misaligned instances decreased across the board, most significantly for RISE (from 10 down to 5) and Sobol’ (from 4 down to 1). This tighter focus indicates the model became more effective at ignoring background noise, thereby correctly identifying benign cases and reducing false positives. This mechanistic improvement tracks directly with the increase in Precision from 0.8096 to 0.853.

\section{Limitations}

While this study provides significant evidence linking lesion attention to diagnostic performance, several limitations warrant consideration.

First, the qualitative assessment of attention alignment relied on visual inspection and predefined categories (Fully Aligned, Partially Aligned, Misaligned). Although consistent criteria were applied, this process remains inherently subjective compared to quantitative metrics such as Intersection over Union (IoU) or pixel-level sensitivity scores. Future work could validate these findings using more granular, automated alignment metrics.

Second, the interpretability analysis was restricted to a stratified sample of 40 images (20 melanoma and 20 non-melanoma). This sample size was constrained by the manual nature of the qualitative assessment, which required detailed human review to categorize the alignment of attention maps across multiple explainability methods. While the sample was balanced to ensure fair representation, it may not fully capture the variance in model behavior across the entire validation dataset.

Third, the failure of the YOLO architecture to achieve a competitive F1-score highlights a limitation in applying pure object detection frameworks to this domain. As noted in the results, YOLO’s optimization for spatial localization likely compromised its ability to learn the fine-grained textural features required for melanoma differentiation. This suggests that while localization is important, it cannot replace the textural feature extraction capabilities of standard CNN classifiers.

Finally, despite the use of oversampling to mitigate class imbalance, the underlying datasets (HAM10000 and ISIC-2019) remain heavily skewed toward non-melanoma cases. This inherent imbalance poses a persistent challenge for training robust diagnostic models and may influence the baseline probability of false negatives in clinical deployment scenarios.

\section{Conclusion}

In this study, we investigated the relationship between model attention to lesion areas and the diagnostic performance of melanoma classification models. By leveraging experiments involving bounding box detection, masked image datasets, and transfer learning, we observed that models with increased attention to lesion areas demonstrated a clear correlation with improved diagnostic accuracy. The combined dataset model, in particular, achieved higher alignment of explainability maps with lesion areas while simultaneously outperforming the baseline model across all key metrics.

These results suggest that enforcing focus on clinically relevant features is a viable pathway to enhance diagnostic reliability. Beyond mere performance metrics, our findings highlight the critical role of explainability methods in guiding model development. As AI becomes increasingly integrated into clinical workflows, ensuring that models rely on interpretable features, such as the lesion itself rather than background noise, is essential for building trust. Ultimately, this study underscores the necessity of attention-focused training approaches and demonstrates how interpretability tools can serve not just as auditors, but as active components in the design of trustworthy medical AI.

\section*{Acknowledgments} I would like to thank Dr. Ivan Felipe Rodriguez for his valuable mentorship, guidance on experimental design, and feedback during the preparation of this manuscript.

\bibliographystyle{unsrt}
\bibliography{references}

\end{document}